# Investigation of slippery behaviour of lubricating fluid coated smooth hydrophilic surfaces

Reeta Pant[a], Pritam Kumar Roy[a], Arun Kumar Nagarajan[b] and Krishnacharya Khare[a*]

[a]Department of Physics, Indian Institute of Technology Kanpur, Kanpur - 208016, India

[b]Hindustan Unilever Research Centre, Bangalore - 560066, India

* kcharya@iitk.ac.in

## Abstract

In the recent years many research groups have studied slippery properties on lubricating fluid infused rough surfaces using hydrophobic substrates. These surfaces show excellent slippery behaviour for water and other liquids. Here we demonstrate a simple method to fabricate stable slippery surfaces based on silicone oil coated hydrophilic samples. At room temperature, as prepared samples exhibit non-slippery behaviour due to sinking of water drops inside silicone oil layer because of inherently hydrophilic silicon substrate. Subsequent annealing at higher temperatures provides covalent bonding of silicone molecules at silicon surface making the surface hydrophobic which was confirmed by lubricant wash tests. So the silicone oil coated annealed samples show excellent water repellency, very low contact angle hysteresis and very good slippery behavior. But these surfaces show poor oil stability against drops flow due to cloaking of the oil around water drops which can be prevented by using drops of larger volume or continuous flow of water.

**Introduction**

Synthetically fabricated liquid repellent surfaces are inspired from lotus leaf effect and are very important for understanding various scientific phenomena and several technological applications such as biomedical devices, surface coatings and liquid transportation etc.[1-8] These surfaces have a presence of dual scale roughness and the large amount of air trapped in the rough areas is responsible for the liquid repellent behaviour.[9, 10] However, these surfaces fail to show liquid repellent behaviour for low surface tension hydrocarbons because in these cases trapped air pressure is not sufficient enough to repel these liquid.[11-13] Also these artificially fabricated surfaces have lot of surface inhomogeneities which act as pinning sites for liquid drops. As a consequence, large contact angle hysteresis is also observed in these surfaces.[14] The superhydrophobicity of these surfaces can be permanently damaged under external pressure and vibrations and hence lose their liquid repelling property.[15-17] To avoid all these drawbacks a novel approach of liquid repelling surfaces based on lubricating fluid infused surfaces was introduced by few research groups.[18-24] The inspiration for these lubricant infused slippery surfaces was taken from nature in form of Nepenthes pitcher plants which have aqueous lubricant layer in side walls which act as slippery interface for insects.[25-28] Instead of relying on the force applied by trapped air in the hierarchical surface texture to repel liquids, some amount of lubricating fluid was infused in some porous structures or membranes to make surface slippery.[29-32] A liquid film is defect free and very smooth in nature, so provides a slippery interface between solid substrate and test liquid (which is to be repelled). Introduction of a lubricating film reduces the friction between the test liquid and substrate surface which allows the test liquid to slip easily.[33] The lubricating film also reduces the amount of pinning sites which results in the drastic lowering of contact angle hysteresis.[23, 34-37] A good interaction between lubricant and substrate as well as presence of lubricant on the surface in a proper amount is necessary condition for the substrate to show best slippery behavior.[19] For complete wetting of the substrate with the lubricant, surface energies of both should match well with each other.[19, 23] The second and very important condition is that the lubricating fluid and the test liquid should not be miscible with each other. In the previously reported articles, it is also mentioned that the wettability of the substrate should be higher for the lubricating fluid as compared to that for the test liquid. So for water as a test liquid, the substrate should be hydrophobic in nature for the better adhesion of lubricant with the substrate. A requirement of surface roughness is also advantageous to hold the sufficient amount of the lubricating fluid.[38] These surfaces are highly slippery in nature for liquids with any surface tension.[39] These can also be used for various other technological applications such as self-cleaning, liquid transportations etc. It was observed that these slippery surfaces have a very low sliding angle (tilt angle of the sample at which test liquid drop starts slipping). Another advantage of these surfaces over lotus leaf effect inspired superhydrophobic surfaces is their self-healing property after a physical damage. Because of the flow properties of the lubricating fluid, it again fills the damaged area and the surface recovers its slippery behavior. The pressure stability of these surfaces is also very high as compared to superhydrophobic surfaces.

Although these studies show complete demonstration of various aspects of the lubricating film infused textured surfaces but much work has not been done on lubricating film modified smooth superhydrophilic surfaces. It's a big challenge to fabricate a slippery surface using a superhydrophilic substrate as it is difficult to get a stably adhered lubricating film on these substrates. Also these surfaces have a natural affinity towards water as water contact angle on smooth Si substrate is ~ 10°. Silicone oil also spreads nicely on Si and shows a contact angle ~ 0°. It is reported in previous studies that silicone oil molecules form Si-O-Si covalent bonds with the Si substrates and these bonds start forming at room temperature but this process fastens on annealing.[40]

Here in the present study we demonstrate a novel technique to fabricate samples with stably adhered film of silicone oil using smooth Si as substrate. Annealing the lubricant coated samples at an appropriate temperature provides the stable uniform, chemically homogeneous and defect free silicone oil layer which results in the excellent slippery behaviour. Annealing the samples alters the inherent hydrophilic nature of the sample. The surface energy of the samples get modified after annealing and applying energy minimization conditions the stable configurations were obtained. The critical sliding angle was ~ 4° for these samples. The observed contact angle hysteresis on these surfaces is very low (< 1°). Later on, confirmation for the bonding of silicone oil molecules with the Si substrates was done by ultrasonicating in surfactant.

**Experimental section**

Polished silicon (Si) wafers cut into 2cm×2cm pieces were used as substrates for slippery surfaces. These substrates were cleaned by ultrasonicating in ethanol, acetone and toluene for 5 mins each followed by oxygen plasma cleaning for 30 sec. Silicone oil (Sigma Aldrich, kinematic viscosity ~ 370 cSt) was used as the lubricating fluid which was spin casted on to the cleaned Si substrates. Subsequently the lubricating fluid coated samples were annealed at different temperature and time to optimize the slippery behavior as well as the stability of the lubricating fluid. Wettability of the lubricant fluid coated slippery surfaces was measured with an optical contact angle goniometer (OCA35, DataPhysics Germany) using sessile drop method. 10 μl volume drops of deionized (DI) water were used as a test liquid to investigate the slippery behavior of the fabricated slippery surfaces. Slippery behavior was quantified by measuring water drop velocity on 10° tilted slippery surfaces and the tilt angle was kept constant for all slippery measurements. To investigate the stability of lubricating fluid, 20 ml volume step of water was dispensed drop-wise on a slippery surface, followed by measuring slip velocity of 10 μl volume water drop at the same spot repeatedly. Later we also analyzed the effect of dispensed drop size (volume) on the stability behavior.

**Results and Discussions**

Cleaned silicon substrates, having a thin native $SiO_2$ top layer, are inherently hydrophilic as well as oleophilic with water and oil contact angles as 10° and 0° respectively. After spin coating thin layer ofsilicone oil, which acts as lubricating fluid, water drops show hydrophobic behavior with equilibrium contact angle as 110°. But these water drops, on room temperature, on such surfaces are very unstable as they sink in the lubricating fluid after some time showing transition from Young's wetting to Neumann's wetting.(ref)Supplementary movie S1 shows the sinking behaviour of a water drop on non-annealed silicone oil coated Si substrate. In the present case, the sinking behavior is huge as the solid surface is inherently hydrophilic. All previous reports on lubricant coated slippery surfaces didn't notice this sinking behavior as they used hydrophobic substrates.[23] Due to the poor adhesion of silicone oil on Si substrates, water drop replaces silicone oil molecules and comes more into contact with Si substrate. Polar interaction between Si and water molecules results in the sticking of water drop to Si substrate and consequently the non-slippery behaviour of the slippery surface. So the most important requirement in using silicon substrates as slippery surfaces is to make them hydrophobic first. This was achieved by annealing the lubricating fluid coated substrates at different temperatures for different time. Aizenberg *et al.* investigated that the wettability of lubricating fluid should be much higher (oleophilic) and test liquid should be non-wetting on the substrate (hydrophobic).[23] In the present case (water and silicone oil as a test liquid and lubricating fluid respectively), we calculated the total energy for the three different configurations: configuration 1 for water on Si substrate, configuration 2 for oil on Si substrate and configuration 3 for water on silicone oil on Si substrate (*cf.* Figure 1). Total energy of the configurations are calculated in terms of various surface and interfacial energies where $\gamma_w$, $\gamma_o$ are surface tensions of water, silicone oil and and $\gamma_{sw}$, $\gamma_{so}$ and $\gamma_{wo}$ are interfacial tensions between solid - water, solid - oil and water - oil respectively. $\theta_o$ and $\theta_w$ represent silicone oil and water contact angles on Si substrate. Condition for slippery surface which can be written in terms of the energy difference between configurations 2 and 3 with that of configuration 1 as follows:

$$\Delta E_2 = E_1 - E_2 = (\gamma_o \cos\theta_o - \gamma_w \cos\theta_w) + (\gamma_w - \gamma_o) \tag{1}$$

$$\Delta E_3 = E_1 - E_3 = (\gamma_o \cos\theta_o - \gamma_w \cos\theta_w) - \gamma_{wo} \tag{2}$$

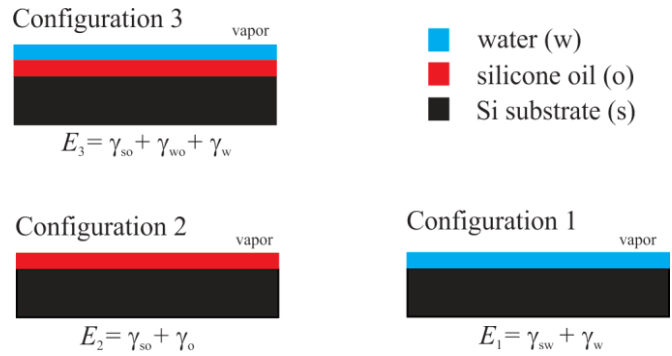

**Figure1.** Schematic diagram of configurations 1, 2 and 3 to calculate total energy of the system. Configurations 1, 2 and 3 correspond to water on substrate, oil on substrate and water on oil on substrate respectively.

To achieve stable slippery surface the energy difference $\Delta E_2$ and $\Delta E_3$ should be positive. For non-annealed samples the value of these energy differences are $\Delta E_2 = 1.1$ mJ/m$^2$ and $\Delta E_3 = -92.5$ mJ/m$^2$. As a result test water drops feel larger interaction with underlying Si substrate and sink into the lubricating layer destroying the slippery behavior. Upon annealing silicone oil coated Si substrates, silicone molecules get covalently bonded to Si surface modifying $\gamma_s$ and $\gamma_{sw}$.[40] Therefore for the annealed samples the energy differences become positive as $\Delta E_2$ = 94.3 mJ/m$^2$ and $\Delta E_3$ = 0.65 mJ/m$^2$ thus preventing test water drops from sinking and making the slippery surface stable.

Subsequently, the effect of annealing temperature and time was investigated to optimize these annealing parameters. Figure 2 shows the effect of annealing temperature and time the slip velocity of water drops on the fabricated slippery surfaces. Lubricating fluid coated Si substrates were annealed at different temperatures for 90 mins and it was found that as the annealing temperature increased the slippery behavior also improved as shown in Fig. 2(a). No improvement in slippery behavior could be observed from room temperature till 50°C and further increasing the annealing temperature up to 200°C makes the samples slippery. Sample annealed at 150°C shows highest slip velocity and stable water contact angle of 108° making the Si surface hydrophobic (supplementary movie S2). Upon annealing at temperatures beyond 200°C, slippery behavior of the samples get destroyed due to breaking of silicone molecules.[41] At temperature around 300°C, the lubricating film starts dewetting (breaking into drops) and it no more remains as uniform film and completely loses its slippery characteristic. Figure 2 (b) shows the optimization of annealing time for samples annealed at 150°C by measuring the slip velocity. As the annealing time is increased, the slip velocity increases making the surface slippery. The highest slip velocity is obtained for the sample annealed for 90 mins which provides enough time to silicone molecules to covalently graft with Si surface. Therefore one could obtain the most efficient silicon oil coated Si substrate based slippery surfaces upon annealing them at 150°C for 90 mins. Figure 2 also predicts that varying thickness of silicone oil (prepared by changing spin coating *rpm* from 2000 up to 8000) does not affect the slippery behavior. This is due to the fact that for the annealed samples, the slip velocity is independent of the lubricant thickness in the given thickness range. So the optimized slippery surfaces, for all subsequent experiments, were fabricated by spin coating silicone oil at 8000 rpm and then annealing at 150°C for 90 mins.

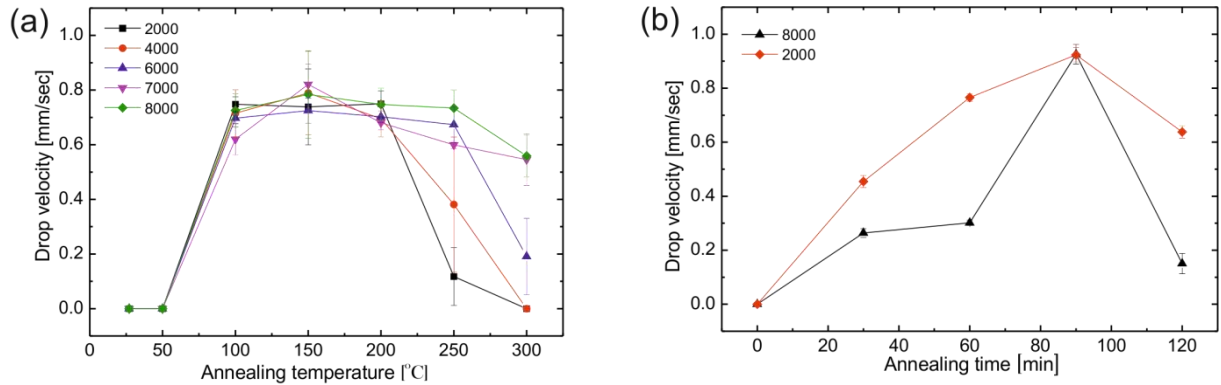

**Figure2.**(a) Slip velocity of a test liquid (water) drop as a function of annealing temperature for lubricating films prepared at different rpm (annealing time was kept constant as 90 min), (b) Optimization of annealing time (annealing temperature was kept constant as 150$^o$C).

To confirm the chemical modification of Si surface upon annealing, which is responsible for stable slippery behavior as suggested by the energy calculations (cf. Figure 1), we slowly removed the lubricant from non-annealed and annealed samples and measured the water contact angle as well as slip velocity. Lubricant coated samples were washed in 2 wt% surfactant (Sodium Dodecyl Sulphate) solution in water via ultrasonicating to remove the lubricant layer slowly and water contact angle and slip velocity were measured as a function of washing time. Ultrasonication in surfactant-water solution slowly removes the silicone oil from the slippery surfaces. For non-annealed samples, the slip velocity remains zero throughout the lubricant removal time. Whereas the water contact angle on non-annealed samples decreases due to thinning of the silicone oil layer and exposure to hydrophilic Si surface. Upon complete removal of lubricant, water contact angle becomes ~ 24° which is very close to the contact angle of bare Si surface. This confirms that the adhesion between silicone molecules and Si surface is very poor and all silicone molecules can be removed via ultrasonication in surfactant solution for few minutes.

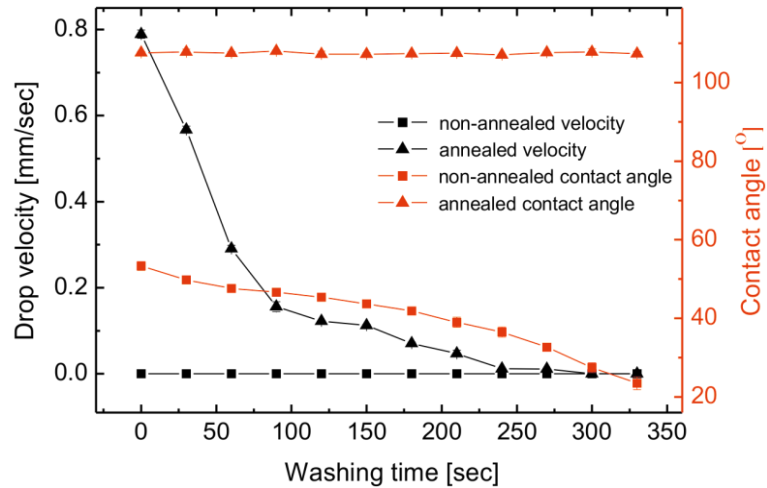

**Figure 3.** Velocities and contact angles of test liquid (water) drops upon removing (by washing) the lubricant layer of non-annealed and annealed slippery surfaces.

On the other hand, for the samples annealed at 150°C for 90 mins, the slip velocity decreases as a function of lubricant washing time. This happens because the surfactant solution slowly removes silicone oil molecules from lubricant layer and its thickness decreases thus decreasing the slip velocity. Upon complete washing, the slip velocity also becomes zero making the surface non-slippery. Initial water contact angle for annealed samples is ~108° which does not vary significantly upon lubricant washing. Even after complete washing, the water contact angle remains 108°. This confirms, for annealed samples, that even after complete washing of lubricant there is at least a monolayer of silicone molecules at Si surface which cannot be removed. This monolayer of silicone molecules can be stable against surfactant washing at Si surface only if they are covalently bonded with Si surface. That is why water contact angle decreases on non-annealed samples but does not vary on the annealed ones. These covalently bonded silicone molecules make the Si surface hydrophobic as well as oleophilic which provides the necessary condition for a stable slippery surface.

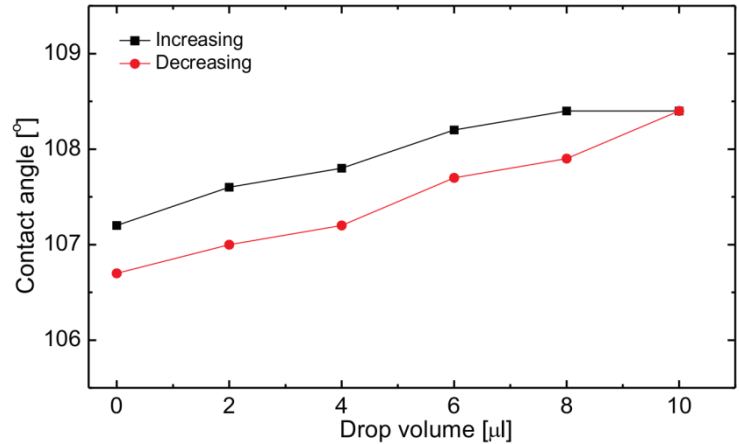

**Figure 4.** Contact angle hysteresis of a water drop on silicone oil coated slippery surface annealed at 150°C for 90 mins.

Contact angle hysteresis measurements were also performed on optimized slippery surfaces. The hysteresis was measured by measuring the advancing and receding contact angles while increasing and decreasing the water drop volume. Figure 4 shows the advancing and receding contact angles and the resulting hysteresis. Due to the slippery characteristic of the surface, three phase contact line of a water drop could move easily upon increasing and decreasing its volume thus resulting in very low hysteresis ($\Delta\theta < 1°$). It confirms that the lubricating fluid surface is homogeneous, defect free and very smooth. Similar very low hysteresis was also observed on samples prepared with different thicknesses of lubricating fluid.

The optimized slippery surfaces show degradation in slip velocity if very large number of drops are dispensed through them. Therefore we also performed the longevity test of the slippery surfaces. For this, slippery behavior was investigated after every step of 20 ml volume of water was dispensed as drops from the slippery surfaces. Contact angle hysteresis was also measured along with slip velocity to predict any change in the surface morphology.

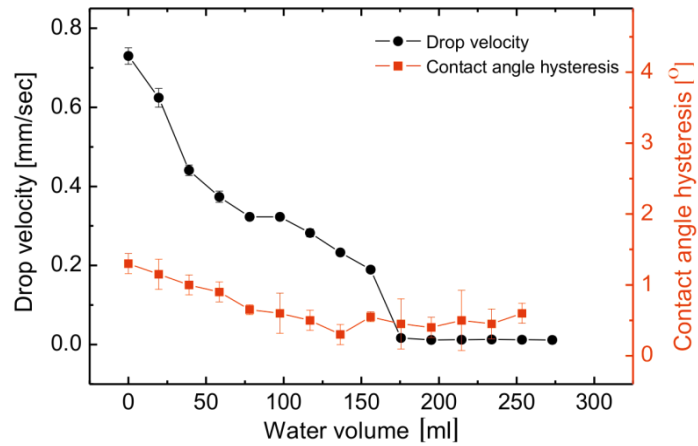

**Figure 5.** Durability test of slippery surfaces by measuring the slip velocity (black data, left Y-axis) and contact angle hysteresis (red data, right Y-axis) as a function of dispensing water.

Slip velocity decreases monotonously after each step which is an indication of the removal of the lubricant layer responsible for slippery behavior. After dispensing about 200 ml of water, the slip velocity becomes zero indicating almost complete removal of lubricant layer. Varanasi et al. have shown that if a water drop is deposited on a silicone oil coated solid surface, a thin layer of silicone oil cloaks the water drop. The condition of the cloaking is defined with positive spreading parameter of oil on water i.e. $S_{ow} = \gamma_w - \gamma_{wo} - \gamma_o$. For our specific case, water and silicone oil, the spreading parameter is found to be 8.5 mN/m which being a positive number ensures the cloaking. Therefore during the dispensing of water drops, silicone oil is also removed continuously resulting in the degradation of the slip velocity. We also checked that this degradation also depends on the volume of the dispensing water drops. The smaller is the dispensing water volume, larger will be degradation of the slippery behavior due to large surface to volume ratio of smaller volume drops. Therefore if dispensing drops are much bigger in volume (say > 100 µl), total degradation of the slippery behavior would take about tens of litres of water. Similarly for continuous stream of water, since cloaking cannot occur, the slippery surface will be stable for hundreds of litres of water.

In case if the slippery behavior is completely destroyed either due to flow or some other reason, it can be quickly recovered upon coating the lubricant again. Since the slippery samples have already been annealed once, there won't be any need for annealing. We also checked that these surfaces behave as slippery not only for water but also for other silicone oil immiscible test liquids e.g. glycerol, ethanol and ethylene glycol.

Conclusions

Hydrophilic Si surfaces coated with silicone oil as lubricating fluid depicts very poor slippery behavior as slipping water drops sink inside the lubricating layer immediately. To prevent the sinking of water drops, the Si surface has to be hydrophobic as well as oleophilic which could be obtained by annealing the silicone oil coated Si surfaces. Annealing at 150°C for 90 mins provides the optimized slippery surface in terms of slip velocity, contact angle and contact angle hysteresis. Hydrophobic modification of Si surface upon annealing was confirmed by washing the lubricating layer in a surfactant solution which systematically decreased the slip velocity but didn't affect the contact angle. Later, the slippery surfaces were found to be degrading as water drops slip on them. This degradation is due to cloaking of silicone oil around water drops.

Acknowledgement

This research work was supported by Hindustan Unilever Limited, India and DST, New Delhi through its Unit of Excellence on Soft Nanofabrication at IIT Kanpur.